\documentclass[]{spie}
\usepackage[]{graphicx}

\title{Enabling Arbitrary Wavelength Optical Frequency Combs on Chip}

\author{
Mohammad Soltani$^1$, Andrey B. Matsko$^2$, and Lute Maleki$^2$
\skiplinehalf $^1$Raytheon BBN Technologies, Cambridge, MA 02138, USA
\skiplinehalf $^2$OEwaves Inc., 465 N. Halstead Street, Suite 140, Pasadena, CA 91107, USA}

\authorinfo{Send correspondence to Mohammad Soltani:
msoltani@bbn.com}

\begin{document}

\maketitle

\begin{abstract}
A necessary condition for generation of bright soliton Kerr frequency combs in microresonators is to achieve anomalous group velocity dispersion (GVD) for the resonator modes. This condition is hard to implement in  visible as well as ultraviolet since the majority of optical materials are characterized with large normal GVD in these wavelength regions. We overcome this challenge by borrowing ideas from strongly dispersive coupled systems in solid state physics and optics. We show that photonic compound ring resonators can possess large anomalous GVD at any desirable wavelength, even if each individual resonator is characterized with normal GVD. Based on this concept we design a mode locked frequency comb with thin-film silicon nitride compound ring resonators in the vicinity of Rubidium D$_1$ line (794.6~nm) and propose to use this optical comb as a flywheel for chip-scale optical clocks.
\end{abstract}

\section{Introduction}

Four-wave mixing (FWM) and Kerr frequency comb generation in optical microresonators has attracted a broad spectrum of interests in science and technology because of both classical and quantum functionalities of the phenomena. A partial list of its applications includes wavelength conversion, optical frequency comb, microwave signal generation, chip-scale optical clocks, and generation of non-classical entangled photons (see \cite{kippenberg11s,moss13np} for a review). Achieving anomalous GVD for the resonator modes is a central challenge to attain such functionalities. This becomes particularly difficult at shorter wavelengths wherein  optical materials show normal dispersion as the wavelength of light  approaches optical transitions. The matter strongly interacts with the photons resulting in a reduced propagation speed, and the appearance of normal GVD.

There have been promising efforts to achieve anomalous GVD by optimizing the waveguide cross section dimensions \cite{moss13np,foster06n}, adding appropriate material cladding \cite{riemensberger12oe}, or using slotted waveguide structures \cite{zhang10oe,zhang13ol,bao15josab}. While these techniques enable dispersion engineering, they fail at shorter wavelengths where the material has strong normal GVD.  It is possible to generate dark solitons with normal GVD resonators, but the pulses generally are not transform limited \cite{weiner15np}. A frequency dependence of the resonator attenuation can be utilized to generate short pulses in a normal GVD resonator \cite{huang15prl}. Recent interesting works demonstrate dispersion engineering of weakly coupled resonators, but on a small range and with limited bandwidth \cite{xue15cleo,miller15opex}.

\begin{figure}
\centering 
\includegraphics[width=14cm]{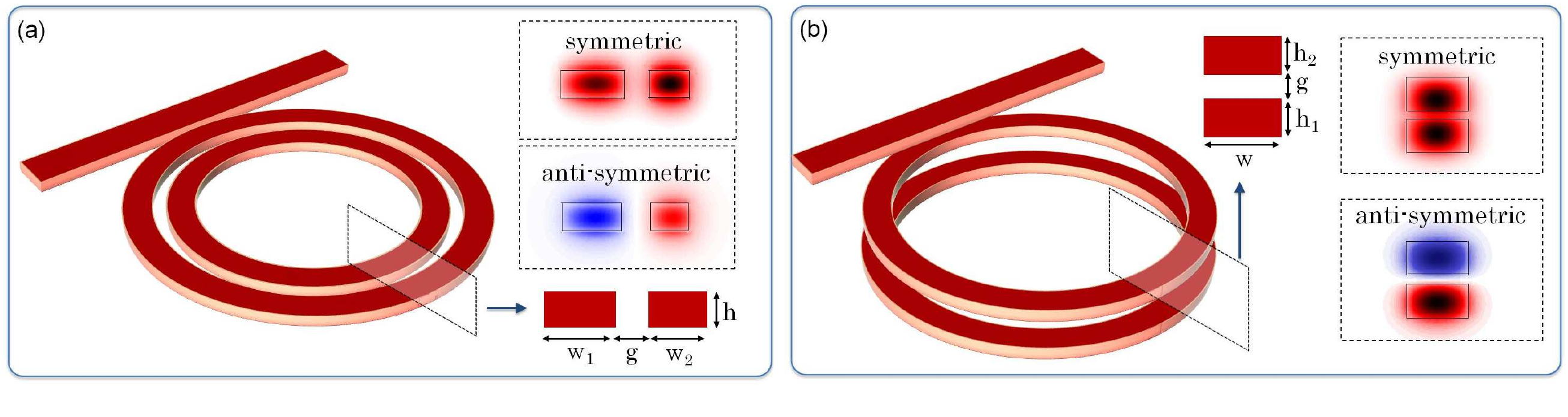}
\caption{{\small Schematic of a dual-ring resonator made of a coupled-waveguide (CWG) structure in (a) lateral and (b) vertical coupling schemes. In each case, the CWG is a directional coupler with the cross section shown in the inset and it supports symmetric and anti-symmetric modes as shown in the insets, respectively. In either structures in (a) and (b), the external coupling waveguide can be integrated laterally or vertically with respect to resonators.}}
\label{fig1}
\end{figure}

In this paper we present a general scheme to achieve anomalous GVD for generating bright solitons at any wavelength in photonic resonators implemented with conventional semiconductor manufacturing techniques. We describe a solution by borrowing ideas from strongly dispersive coupled systems in solid state physics and optics.  It is known that the coupling of electron waves in crystalline solids results in both positive and negative effective electron mass \cite{kittel86book}. The electronic energy band formation in these systems is similar to optical supermode formation in the case of interacting coupled optical waveguides or resonators, and the positive and negative effective electron mass is formally analogous to normal and anomalous GVD of the optical supermodes. We further show that the GVD of two coupled systems can vary over a large range if they are nearly phase-matched but have different group velocities. This coupling scheme, which is strongly dispersive, is observed in polaritons (phonon-photon coupling), where photons and phonons are phase matched but have different group velocities \cite{kittel86book}. Such coupling has been also studied in optical fiber systems to engineer the GVD \cite{peschel95apl}.

We show that the GVD of a selected supermode of a compound microring resonator made of a composite waveguide (Fig.~\ref{fig1}) can be properly adjusted to achieve anomalous dispersion, while each individual waveguide has large normal GVD. The proposed method provides the ability to generate Kerr frequency combs at any desirable wavelength utilizing on-chip planar resonators. Using this concept we design a compound ring resonator with anomalous GVD at 795~nm for generation of bright soliton Kerr frequency combs. For the design and simulation we use silicon nitride (Si$_3$N$_4$) which is a low loss broadband photonic material providing high-Q planar ring resonators ($Q>10^{6}$) with high 3$^{\rm rd}$ order optical nonlinearity \cite{barclay06apl}. For all the analyses the (Si$_3$N$_4$) resonators have silicon oxide (SiO$_2$) cladding. This compound structure allows utilizing thin film Si$_3$N$_4$ platforms, that are robust and stress-free, enabling wafer-scale devices with desirable anomalous GVD.

\section{Group Velocity Dispersion of Supermodes of a Dual-Ring Structure}

Let us consider a compound dual-ring structure created from two strongly coupled dielectric waveguides (CWG) (Fig.~\ref{fig1}). We study two architectures for this system: i) lateral, wherein the rings are concentric and laterally coupled (Fig.~\ref{fig1}a); and ii) vertical, wherein the rings are vertically laid on each other (Fig.~\ref{fig1}b). These structures are different from a conventional slot waveguide structures as the inner spacing between the rings can be quite large that no slot mode is supported. The compound resonators support two supermodes: (1) symmetric and (2) anti-symmetric. Dispersion of the modes can be modified by changing the spatial gap between the rings in addition to their dimensions. Thus the supermodes can have anomalous GVD even if the individual waveguides have normal GVD.

For the sake of simplicity, let us assume an infinite bend radius for the microrings so that we could use the coupled-mode analysis of parallel straight waveguides (later on we present results of accurate numerical simulations taking the bend radius into account). The coupling equations for two identical waveguides can be written as \cite{haus83book}:
\begin{eqnarray} \label{a1}
\frac{d}{dz}a_{1}= i \beta a_1+i \kappa a_2,\\
\frac{d}{dz}a_{2}= i \beta a_2+i \kappa^* a_1, \label{a2}
\end{eqnarray}
where $a_1$ and $a_2$ are the amplitude of the field in each waveguide, $\beta$ is the propagation constant of the mode in individual isolated waveguides, $\kappa$ is the coupling factor between the waveguides, and $z$ is the propagation direction. By finding the eigen solutions of set (\ref{a1}) and (\ref{a2}), the propagation constants of the symmetric ($\beta_+$) and anti-symmetric ($\beta_-$) modes of the CWG structure are obtained as: $\beta_{\pm}=\beta \pm |\kappa|$. Because the dispersion of individual waveguides is $D_0=-(\omega^2/(2\pi c))d^2 \beta /d\omega^2$, the dispersion of the symmetric and anti-symmetric mode can be obtained as:
\begin{equation} \label{disp}
D_{\pm}=D_0 \mp \frac{\omega^2}{2 \pi c} \frac{d^2 |\kappa|}{d\omega^2}
\end{equation}

Since the the coupling between the waveguides in the compound structure depends on the evanescent field, the coupling factor is related to frequency as $|\kappa|\sim \exp⁡(-\alpha \omega)$, wherein $\alpha$ is some positive constant that depends on the material and structure of the waveguides as well as the cladding. The 2$^{\rm nd}$ order derivative of the coupling-originated factor in Eq.~(\ref{disp}) is positive, so that $D_->D_0$. In other words, the coupling shifts the integral GVD of the anti-symmetric supermode towards the anomalous region. In addition, and as shown later, when the two waveguides have different group velocities, their coupling can be strongly dispersive providing a large range anomalous GVD. These are the basic idea utilized in this study.

\begin{figure}[hts]
\centering
\includegraphics[width=10cm]{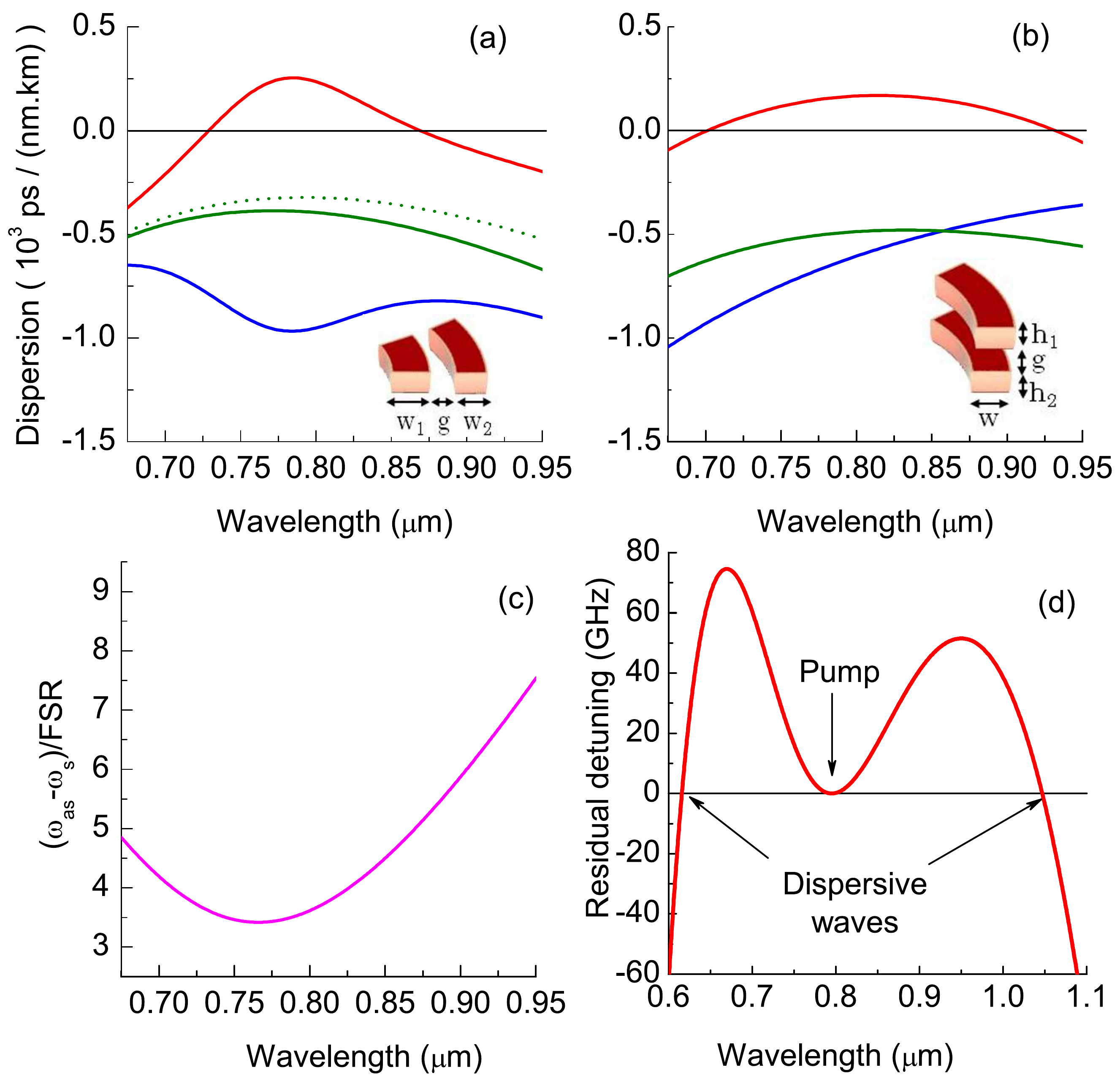}
\caption{{\small Achieving anomalous dispersion in the visible range with the anti-symmetric mode (red plot) of a compound Si$_3$N$_4$ ring resonator for (a) lateral and (b) vertical designs. The polarizations in (a) and (b) are transverse magnetic (TM, electric field has out-of-plane direction) and transverse electric (TE, electric field has in-plane direction), respectively. The insets in (a) and (b) show portion of the dual-rings.  The results for the symmetric mode (blue plot) and the mode of a single ring (green plot) are also shown. In (a) dotted (solid) green line stands for the external (internal) ring. For (a) the external ring radius is 100 $\mu$m, $w_1 = 650$~nm, $w_2 = 550$~nm, $h = 450$~nm, and $g = 300$~nm. For (b) the ring radius is 100 $\mu$m, $w = 700$~nm, $h_1 = h_2 = 350$~nm, and $g = 100$~nm. Red arrow illustrates the wavelength at which generation of Kerr frequency comb was simulated. (c) Variation of resonance frequency difference between the symmetric and antisymmetric modes normalized to FSR of the single ring vs. the resonance wavelength of the single ring. (d) Residual frequency detuning characterizing dispersion of the anti-symmetric supermode spectrum defined as $\omega_j-\omega_{j_0}-\omega_{FSR}(j-j_0)$, where $\omega_{j_0}$ is the frequency of the pumped mode having number $j_0$, $\omega_{j}$ is the frequency of a $j^{\rm th}$ mode, $\omega_{FSR}=(\omega_{j_0+1}-\omega_{j_0-1})/2$.}}
\label{fig2}
\end{figure}

To achieve anomalous GVD the rings in the compound structure need to be phase-matched for efficient mutual coupling. Phase matching for the lateral architecture requires the effective refractive index of the external ring to be smaller than that of the internal ring and scales with the inverse of the ring radius. For that we reduce the width of the external ring to fulfill the requirement. This modification also results in different group velocities for individual rings and provides strong dispersive coupling. For vertical architecture (Fig.~\ref{fig1}b) phase-matching is readily achievable as the rings can be made identical.

We perform exact numerical simulations of the lateral and vertical ring structures taking into account bending of the resonators as well as the material dispersion of Si$_3$N$_4$ and SiO$_2$ cladding. Figure~\ref{fig2}a shows dispersion of the symmetric and anti-symmetric supermodes of a laterally-coupled dual-ring resonator designed for visible wavelengths. The resulting GVD is compared with that of a single ring resonator. One can see that the anti-symmetric mode provides anomalous GVD over a broad wavelength range, while the single ring naturally has normal GVD. The dispersion of symmetric and anti-symmetric supermodes of the dual-ring in Fig.~\ref{fig2}a shows a peaks at the wavelength of the maximum phase matching. The magnitude of this peak depends on the group velocity difference between the modes of the individual rings.

Figure~\ref{fig2}b shows the simulation results for the vertically coupled compound resonator. The resultant anomalous GVD of the anti-symmetric supermode is less frequency dependent than that of the lateral architecture, since both rings are identical and equiradial. The fabrication of vertical dual-ring is more involved, as it requires multi-layer deposition of materials, but since the thickness of each ring is quite small (350~nm for Fig.~\ref{fig2}b), the implementation of this structure is feasible with the existing and mature multi-layer integration \cite{li13oe}.

The compound resonators discussed here are based on strong coupling between the individual rings. Figure~\ref{fig2}c illustrates the frequency split between symmetric and anti-symmetric modes of the same order normalized to the free spectral range (FSR) of the compound resonator with data in Fig.~\ref{fig2}b. This splitting exceeds the FSR, which shows that these supermodes are independent. This contrasts with the case of weakly coupled rings where the two supermodes are correlated with eigen-modes of each ring.

In the strongly coupled case considered here, symmetric and anti-symmetric supermodes have slightly different FSRs and significantly different GVD. This property can be used for observation of various GVD-sensitive dissimilar nonlinear processes in the same compound structure. Switching between these processes can be achieved by selection of the proper supermode family. We illustrate this for the case of Kerr frequency comb generation.

For the rest of the discussion we only focus on the lateral coupling architecture. The GVD illustrated by Fig.~\ref{fig2}a allows generating an octave spanning Kerr frequency comb. To show that, we need to depict higher order dispersion terms, as illustrated by Fig.~\ref{fig2}d. The spectral width of the comb is approximately limited by the region confined between the two points corresponding to the generation of Cherenkov radiation \cite{brasch14arch}. Such a broad comb can be stabilized using f-2f or 2f-3f techniques \cite{udem02n} making it suitable for clocks and oscillators.

\section{Generation of Dark and Bright Solitons from the Same Microring}

\begin{figure*}
\centering
\includegraphics[width=14cm]{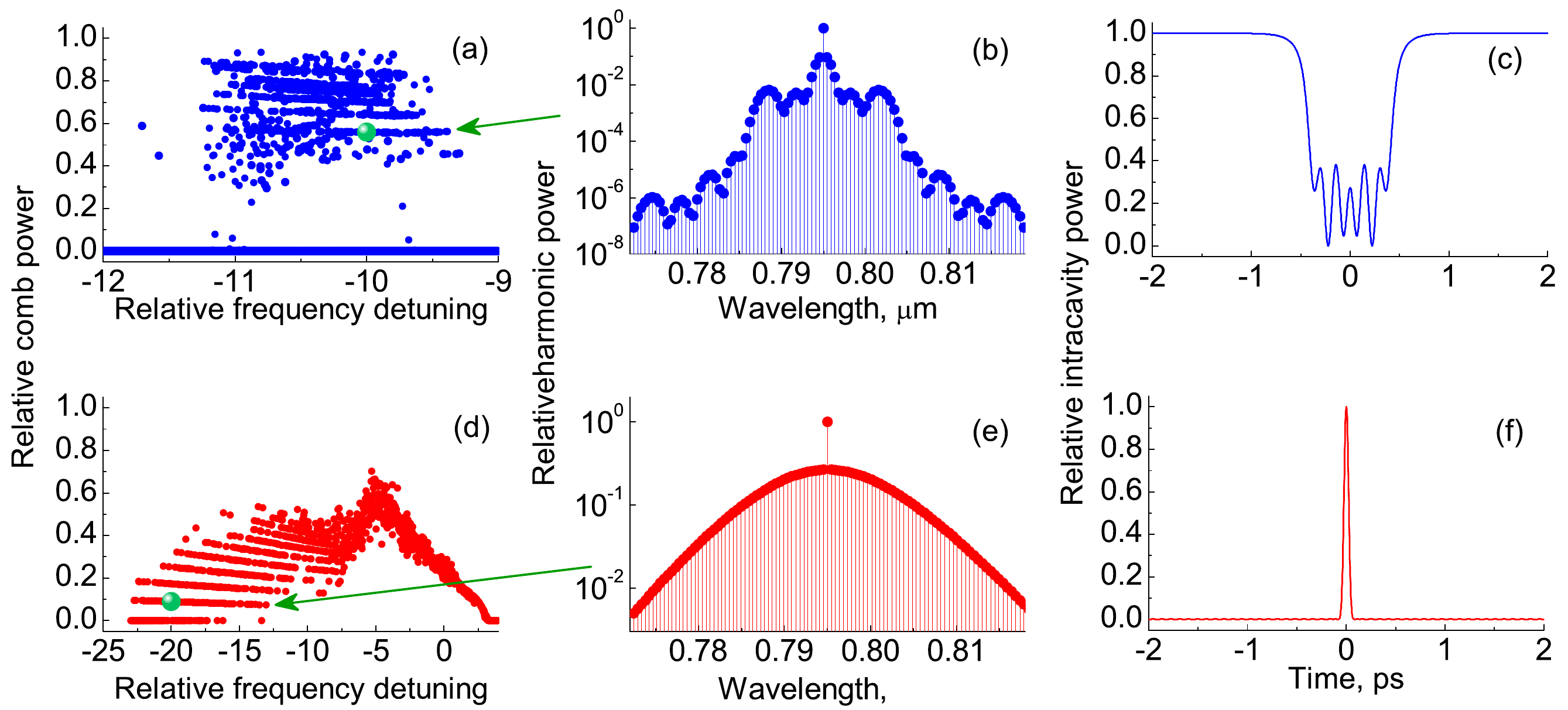}
\caption{ {\small Frequency comb generation regimes in the case of normal (panels (a), (b), and (c)) and anomalous (panels (d), (e), and (f)) dispersion observed for symmetric ($FSR = 219$~GHz) and antisymmetric ($FSR = 218$~GHz) supermodes of the resonator with dispersion parameters illustrated in Fig.~\ref{fig2}a.   Isolated branches in (a) and (d) corresponding to soliton generation within the resonators are clearly visible. Panels (a) and (d) stand for numerical simulation of the power of the comb sidebands versus detuning of the pump light frequency from the corresponding mode of the resonator. The detuning is normalized to the bandwidth of the optical modes. The total power of the sidebands is normalized to the input pump power. Panels (d) and (e) illustrate fundamental frequency combs generated within the compound resonator structure at the frequency detuning marked by green dots in (a) and (d).  The power of the comb harmonics is normalized to the pump power within the resonator mode. The fundamental frequency combs correspond to a single pulse confined in the resonator which are dark, (c), and bright (f), solitons. The power of the bright (dark) solitons is normalized to the peak (background) power. For the simulations, the pump wavelength was selected to be at 795~nm.}}
\label{fig3}
\end{figure*}

We numerically study \cite{chembo10pra} Kerr frequency comb generation regimes for the structure in Fig.~\ref{fig2}a when the pump light is in resonance with either of the supermode families near 795~nm. Figures~(\ref{fig3}a-c) and (\ref{fig3}d-f) characterize the comb generation for symmetric and anti-symmetric supermodes, respectively. Multiple branches corresponding to generation of mode locked frequency combs are observed in both (\ref{fig3}a) and (\ref{fig3}d). The lowest branches correspond to generation of a frequency comb with single pulse soliton within the resonator \cite{matsko12pra,herr14np}. These pulses and corresponding frequency combs can be seen at other panels of Fig.~\ref{fig3}.

From the results of the numerical simulations we conclude that while frequency combs can be generated at both normal and anomalous regimes, the anomalous regime is more advantageous since i) the dynamic range of the frequency comb generation is much larger, ii) the mode locked regime is much better isolated from the chaotic and breathing regimes, iii) the envelope of the comb is smoother, and iv) the bright pulse generated in the resonator can be very short, approaching one femtosecond range. Therefore, the method described in this paper is truly advantageous for Kerr frequency comb generation applications, including chip scale optical clocks \cite{savchenkov13ol,papp14o}. In this clock architecture a frequency harmonic of the comb is locked to optical Rb transition and the comb transfers the optical stability to microwave frequency corresponding to its repetition rate.

By varying the parameters of the Si$_3$N$_4$  compound ring we were able to achieve anomalous GVD at any wavelength within 450~nm--2500~nm range. The method is also applicable to any other platform, including crystalline optical materials such as CaF$_2$, MgF$_2$, Al$_2$O$_3$, and others. This means that the technique can be used for generation of frequency combs at virtually any desirable wavelength, though further study is required to verify any impact of multi-photon absorption \cite{moss13np} as well as change of the coupling strength \cite{huang15prl} on the efficiency of the comb generation.

\section{Conclusion}

We have presented a generalized dispersion engineering techniques based on  strong coupling of compound ring resonators to achieve anomalous GVD.  This approach is suitable for nonlinear optics applications and generation of mode-locked frequency combs at arbitrary wavelength on a chip. To validate our technique, we numerically demonstrated stable frequency combs and cavity solitons at wavelengthes near the transition of Rubidium. We expect that this dispersion engineering method and the proposed compound dual-resonator architecture will find wide applications in many areas including chip scale nonlinear quantum photonics, metrology and spectroscopy.

\section*{Acknowledgement}
We gratefully acknowledge Dr. Kartik Srinivasan (NIST) for providing the data on Si$_3$N$_4$ refractive index. A.M. and L.M. acknowledge partial support from DARPA contracts W911QX-13-C-0141 and HR0011-15-C-0054.


\begin{thebibliography}{00}

\bibitem{kippenberg11s} T. Kippenberg, R. Holzwarth, and S. Diddams, Science {\bf 332}, 555 (2011).

\bibitem{moss13np} J. Moss, R. Morandotti, A. L. Gaeta, and M. Lipson, Nature Photonics {\bf 7}, 597  (2013).

\bibitem{foster06n} M. A. Foster, A. C. Turner, J. E. Sharping, B. S. Schmidt, Mi. Lipson, and A. Gaeta, Nature {\bf 441}, 960 (2006).

\bibitem{riemensberger12oe} J. Riemensberger, K. Hartinger, T. Herr, V. Brasch, R. Holzwarth, T. J. Kippenberg, Opt. Express {\bf 20}, 27661 (2012).

\bibitem{zhang10oe} L. Zhang, Y. Yue, R. G. Beausoleil, and A. E. Willner,  Opt. Express {\bf 18},  20529 (2010).

\bibitem{zhang13ol} L. Zhang, C. Bao, V. Singh, J. Mu, C. Yang, A. M. Agarwal, L. C. Kimerling, and J. Michel,  Opt. Lett. {\bf 38}, 5122 (2013).

\bibitem{bao15josab} C. Bao, Y. Yan, L. Zhang, Y. Yue, N. Ahmed, A. M. Agarwal, L. C.Kimerling, J. Michel, and A. E. Willner, J. Opt. Soc. Am. B {\bf 32}, 26 (2015).

\bibitem{weiner15np} X. Xue, Y. Xuan, Y. Liu, P.-H. Wang, S. Chen, D. E. Leaird, M. Qi, and A. M. Weiner, Nature Photonics, {\bf 9}, 594 (2015).

\bibitem{huang15prl} S.-W. Huang, H. Zhou, J. Yang, J.-F. McMillan, A. Matsko, M. Yu, D.-L. Kwong, L. Maleki, and C.-W. Wong,  Phys. Rev. Lett. {\bf 114}, 053901 (2015).

\bibitem{xue15cleo} X. Xue, Y. Xuan, P.-H. Wang, Y. Liu, D. E. Leaird, M. Qi, and A. M. Weiner, $arXiv:1503.06142v1$ (2014).

\bibitem{miller15opex} S. Miller, Y. Okawachi, S. Ramelow, K. Luke, A. Dutt, A. Farsi, A. L. Gaeta, and M. Lipson, Opt. Express 23, 21527 (2015).

\bibitem{kittel86book} C. Kittel, {\em Introduction to Solid State Physics,} (Wiley, 1986).

\bibitem{peschel95apl} U. Peschel, T. Peschel, and F. Lederer,  Appl. Phys. Lett. {\bf 67}, 2111 (1995).

\bibitem{barclay06apl} P. E. Barclay, K. Srinivasan, O. Painter, B. Lev, and H. Mabuchi, App. Phys. Lett. {\bf 89}, 131108 (2006).

\bibitem{haus83book}	H. Haus, {\em Wave and fields in optoelectronics,} (Prentice Hall, 1983).

\bibitem{li13oe} Q. Li, A. A. Eftekhar, M. Sodagar, Z. Xia, A. H. Atabaki, and A. Adibi, Opt. Express {\bf 21}, 18236 (2013).

\bibitem{brasch14arch} V. Brasch, T. Herr, M. Geiselmann, G. Lihachev, M. H. P. Pfeiffer, M. L. Gorodetsky, and T.J. Kippenberg, $arXiv:1410.8598$ (2014).

\bibitem{udem02n} T. Udem, R. Holzwarth, and T. W. Hansch, Nature  {\bf 416}, 233 (2002).

\bibitem{chembo10pra} Y. K. Chembo and N. Yu,  Phys. Rev. A {\bf 82}, 033801 (2010).

\bibitem{matsko12pra} A. B. Matsko, A. A. Savchenkov, V. S. Ilchenko, D. Seidel, and L. Maleki, Phys. Rev. A {\bf 85}, 023830 (2012).

\bibitem{herr14np} T. Herr, V. Brasch, J. D. Jost, C. Y. Wang, N. M. Kondratiev, M. L. Gorodetsky and T. J. Kippenberg,  Nature Photonics {\bf 8}, 145 (2014).

\bibitem{savchenkov13ol} A. A. Savchenkov, D. Eliyahu, W. Liang, V. S. Ilchenko, J. Byrd, A. B. Matsko, D. Seidel, and L. Maleki, Opt. Lett. {\bf 38}, 2636 (2013).

\bibitem{papp14o} S. B. Papp, K. Beha, P. Del’Haye, F. Quinlan, H. Lee, K. J. Vahala, and S. A. Diddams, Optica {\bf 7} 10 (2014).


\end{thebibliography}
\end{document}